\documentclass[twocolumn,twoside,slac_two]{revtex4}
\usepackage{graphicx}
\usepackage{fancyhdr}
\usepackage{hyperref}
\hypersetup{
    colorlinks=true,
    urlcolor=magenta,
}
\begin{document}

\title{Precision Measurement of the Proton Flux in 
Primary Cosmic Rays from 1 GV to 1.8 TV with the 
Alpha Magnetic Spectrometer on the International Space Station}

%

\author{Cristina Consolandi on Behalf of the AMS Collaboration}
\affiliation{University of Hawaii at Manoa, 2505 Correa Road, Honolulu 96822, Hawaii, US}

\begin{abstract}
A precision measurement of the proton flux in primary cosmic rays with rigidity from 1 GV to 1.8
TV is presented based on 300 million events. The results show that the proton flux is smooth
and exhibits no sharp structures with rigidity. The detailed variation with rigidity of the flux
spectral index is presented. The spectral index is progressively hardening at high rigidities.
Solar effects are also discussed. 
\end{abstract}

\maketitle

\thispagestyle{fancy}

Protons are the most abundant particles in cosmic rays.
Knowledge of the precise behavior of the proton flux as a function of energy is important 
in understanding the origin, acceleration, and propagation of cosmic rays.
The precise measurement of the proton flux in primary cosmic rays
in the rigidity range from 1\,GV to 1.8\,TV performed by the Alpha Magnetic Spectrometer (AMS)
during the first 30 months of operation onboard the International Space Station (ISS)
is described as published in~\cite{bib:proton}.
Moreover, the first 5 years of data collected from May 2011 to May 2016, reported in this proceeding,
show structures, in the low rigidity range, related to the long and short time scale solar activity.

\section{AMS DETECTOR}
AMS is a general purpose high energy particle physics detector in space.
The layout and description of the detector are presented in~\cite{bib:ams-02}.
The key elements used in this measurement are the permanent magnet, the silicon tracker, 
four planes of time of flight (TOF) scintillation counters, and the array 
of anticoincidence counters (ACC). AMS also contains a transition radiation detector (TRD), 
a ring imaging {\v C}erenkov detector (RICH), and an electromagnetic calorimeter (ECAL).
The central field of the magnet~\cite{bib:newlayout} is 1.4\,kG.
Before flight, the field was measured in 120,000 locations to an accuracy of better than 2\,G. 
On orbit, the magnet temperature varies from $-3$ to $+ 15^\circ$C.
The field strength is corrected with a measured temperature dependence of $-0.09\%/^\circ$C.

The tracker~\cite{bib:tracker} has nine layers, 
the first (L1) at the top of the detector, 
the second (L2) above the magnet,
six (L3 to L8) within the bore of the magnet,
and the last (L9) above the ECAL.
L2 to L8 constitute the inner tracker. 
The tracker accurately determines the trajectory 
of cosmic rays by multiple measurements of the coordinates
with a resolution in each layer of 10$\,\mu$m in the bending  direction.
Together, the tracker and the magnet measure the rigidity $R$ of charged cosmic rays.
The maximum detectable rigidity (MDR) is 2\,TV 
over the 3\,m lever arm from L1 to L9.
Each layer of the tracker also provides an independent measurement 
of the absolute value of the charge $|Z|$ of the cosmic ray.
The charge resolution of the layers of the inner tracker together is 
$\Delta Z \simeq 0.05$ for $|Z| = 1$ particles.
Two planes of TOF counters~\cite{bib:tof} are located above L2 
and two planes are located below the magnet.
The average time resolution of each counter for $|Z| = 1$ particles 
has been measured to be 160\,ps
and the overall velocity ($\beta=v/c$) resolution to be $\Delta\beta/\beta^2 = 4$\%.
This discriminates between upward- and downward-going particles. 
The coincidence of signal from the four TOF planes together with the absence 
of signal from the ACC provides a charged particle trigger. 
The coincidence of 3 out of the 4 TOF layers
with no ACC requirement was instead used to provide an unbiased trigger.
The unbiased trigger, prescaled by 1\%, was used to measure the efficiency
of the charged particle trigger.

\section{SELECTION AND FLUX ANALYSIS}

In the first 30 months ($7.96\times 10^{7}$\,s) AMS collected $4.1\times 10^{10}$ cosmic ray events.
Due to the influence of the geomagnetic field,
this collection time for primary cosmic rays increases with increasing rigidity
becoming constant at $6.29 \times 10^{7}$\,s above 30\,GV.

By selecting events to be downward going and to have a positive 
reconstructed track in the inner tracker with charge compatible with $|Z|=1$, 
$1.1\times 10^{10}$ events were obtained.
In order to have the best resolution at the highest rigidities,
further selections are made by requiring the track to pass through L1 and L9
and to satisfy additional track fitting quality criteria
such as a $\chi^2/d.f.<10$ in the bending coordinate.
In addition, to select only primary cosmic rays, well above the geomagnetic cutoff,
the measured rigidity is required to be greater than 1.2 times the maximum geomagnetic 
cutoff within the AMS field of view.
The cutoff was calculated by backtracing~\cite{bib:back} particles 
from the top of AMS out to 50 Earth's radii
using the IGRF~\cite{bib:IGRF} geomagnetic model.
These procedures resulted in a sample of $3.0\times 10^{8}$ primary cosmic rays with $Z=+1$.

Since protons are the dominant component of cosmic rays, 
the selected sample has only small contributions of other particles, 
mainly deuterons which are not removed in this analysis.
The deuteron contribution decreases with rigidity:
at 1\,GV it is less than 2\% and at 20\,GV it is 0.6\%~\cite{bib:deut-else,bib:deut-ams}.
The sample also contains protons from nuclei which interact at the top of AMS (for example, in L1 or the TRD).
From the measured flux~\cite{bib:ams-he} and Monte Carlo simulation 
this contribution is 0.5\% at 1\,GV decreasing to less than 0.1\% at and above 10\,GV.
Contamination from $e^+$ and $e^-$~\cite{bib:ams-e7p5}, overwhelmingly $e^+$,
was estimated to be  less than 0.1\% over the entire rigidity range.
The background contributions from protons which originated in the interactions of
nuclei at the top of AMS and $e^\pm$, both noticeable only below 2\,GV,
are subtracted from the flux and the uncertainties are accounted 
for in the systematic errors.

After selection, the isotropic proton flux  $\Phi_i$ for the $i^\mathrm{th}$ rigidity bin $(R_i,R_i+\Delta R_i)$ is
\begin{equation}
\Phi_i = \frac{N_i}
               {A_i\, \epsilon_i\, T_i\, \Delta R_i}
\label{eq:flux}
\end{equation}
where 
$N_i$ is the number of events corrected with the rigidity resolution function
by means of the unfolding procedure described in details in~\cite{bib:proton}.
$A_i$ is the effective acceptance,
$\epsilon_i$ is the trigger efficiency,
and $T_i$ is the collection time. 
The proton flux was measured in 72 bins, $i=1$ to 72, from 1\,GV to 1.8\,TV
with bin widths chosen according to the rigidity resolution.
The effective acceptance $A_i$ was calculated using Monte Carlo 
and then corrected for small differences found between the data and Monte Carlo 
event selection efficiencies. 
The trigger efficiency $\epsilon_i$ is measured 
from data with the unbiased trigger events.
The trigger efficiency ranges from 90 to 95\%.
The 5 to 10\% inefficiency is due to secondary $\delta$-rays in the magnetic field entering the ACC.
The Monte Carlo agrees with the measured trigger efficiency within 0.5\%.
The bin-to-bin migration of events was corrected using a rigidity resolution
function obtained from Monte Carlo simulation and verified with test beam data ~\cite{bib:proton}.

Extensive studies were made on systematic errors. 
The errors include uncertainties in the trigger efficiency, 
the acceptance, the background contamination, the geomagnetic cutoff factor, 
the event selection, the unfolding, the rigidity resolution function,
and the absolute rigidity scale.
The trigger efficiency error is dominated by the statistics available from 
the 1\% prescaled unbiased event sample.  
It is negligible (less than 0.1\%) below 500\,GV and reaches 1.5\% at 1.8\,TV.
The geomagnetic cutoff factor was varied from 1.0 to 1.4 and the resulting proton fluxes showed a 
systematic uncertainty of 2\% at 1\,GV and negligible above 2\,GV.
Using the most recent IGRF model~\cite{bib:IGRF12} and the IGRF model with external 
non-symmetric magnetic fields~\cite{bib:tsyg} does not introduce observable changes 
in the flux values nor in the systematic errors.

The effective acceptance was corrected for 
small differences between the data 
and the Monte Carlo samples related to
the event reconstruction
and selection.
Together, the correction was found to be 5\% at 1\,GV decreasing below 2\% above 10\,GV, 
while the corresponding systematic uncertainty is less than 1\% above 2\,GV.

The detector is mostly made of carbon and aluminum.
The corresponding inelastic cross sections of $p+\mathrm{C}$ and $p+\mathrm{Al}$ 
are known to within 10\% at 1\,GV and 4\% at 300\,GV~\cite{bib:cross},
and 7\% at 1.8\,TV from model estimations~\cite{bib:geant}.
{ The inelastic cross sections are used in the
Monte Carlo based estimation of the effective acceptance and,}
to estimate the systematic error due to 
the uncertainty in the inelastic cross sections,
dedicated samples of protons 
were simulated with the $p+\mathrm{C}$ and 
$p+\mathrm{Al}$
cross sections varied by $\pm$10\%.
From the analysis of these samples together with the current knowledge of the cross sections,
a systematic error of 1\% at 1\,GV, 0.6\% from 10 to 300\,GV, and 0.8\% at 1.8\,TV was obtained.

The rigidity resolution function was verified with data from both the ISS and the test beam. 
For this, the residuals between the hit coordinates measured in tracker layers L1 and L9 and
those obtained from the track fit using the information from only the inner tracker L2 to L8
were compared between data and simulation.
In order to validate the alignment of the external layers, the difference 
between the rigidity measured using the information from L1 to L8 and from L2 to L9
was compared between data and simulation.
The resulting uncertainty on the MDR was estimated to be 5\%.
The corresponding unfolding errors were obtained by varying 
the width of the Gaussian core of the resolution function by 5\% 
and the amplitude of the non-Gaussian tails by $\sim$20\% 
over the entire rigidity range and found to be
1\% below 200\,GV and 3\% at 1.8\,TV.

There are two contributions to the systematic uncertainty on the rigidity scale.
The first is due to residual tracker misalignment. 
From the 400\,GeV/\textit{c} test beam data it was measured to be less then 1/300\,TV$^{-1}$. 
For the ISS data this error was estimated by comparing 
the $E/p$ ratio for electron and positron events, 
where $E$ is the energy measured with the ECAL 
and $p$ is the momentum measured with the tracker, see~\cite{bib:ams-pf} for details.
It was found to be $1/26$\,TV$^{-1}$, limited by the current high energy positron statistics.
The second systematic error on the rigidity scale arises from the
magnetic field map measurement 
 (0.25\%)
and its temperature corrections (0.1\%).
Taken in quadrature and weighted by the measured flux,
this amounts to less than 0.5\% 
{systematic error on the flux}
for rigidities above 2\,GV.

\section{RESULTS AND CONCLUSIONS}
\begin{figure}
\includegraphics[width=80mm]{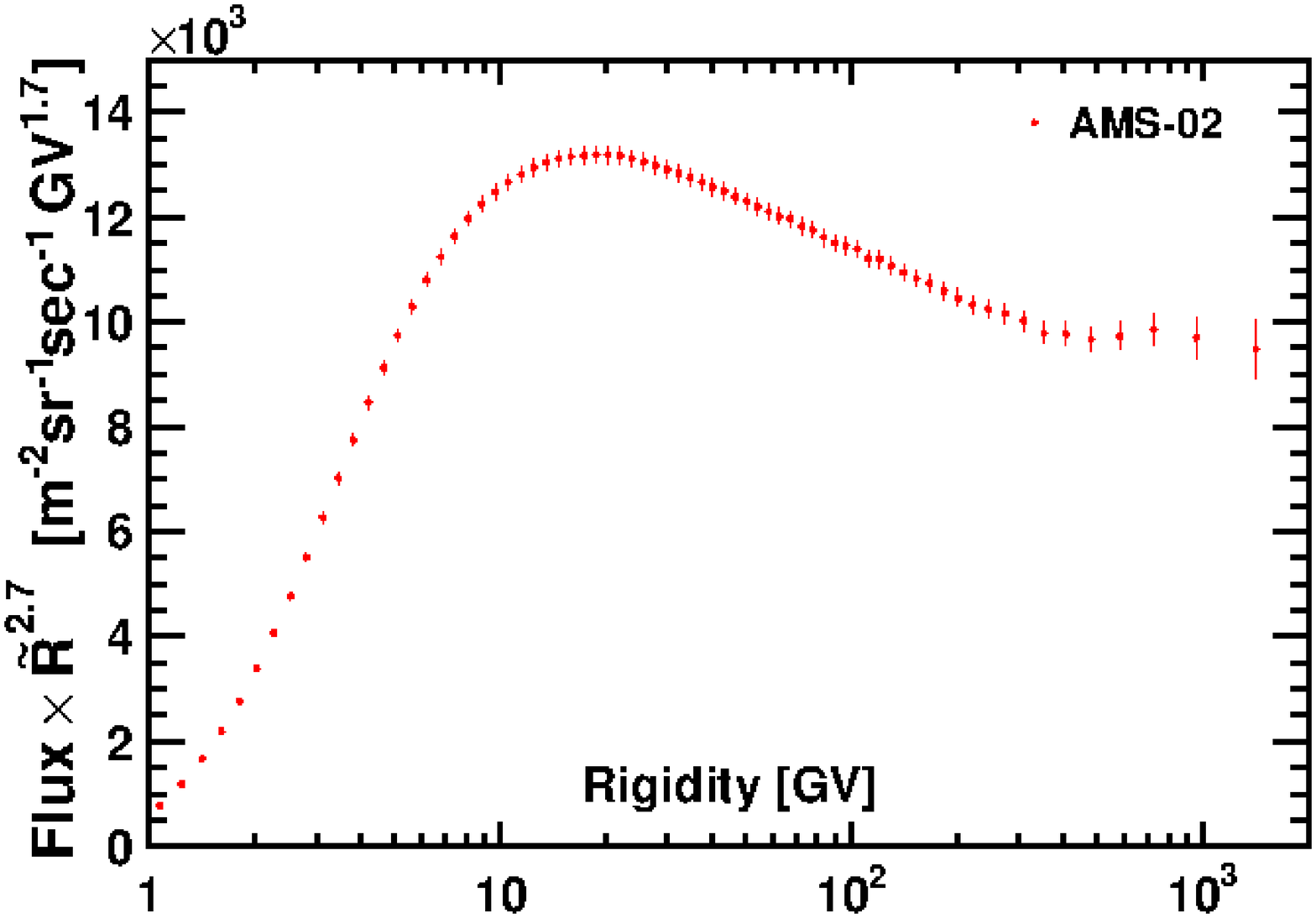} 
\caption{The AMS proton flux multiplied by $\tilde{R}^{2.7}$ 
with its total error as a function of rigidity \label{fig:proton}.}
\end{figure}

\begin{figure}
\includegraphics[width=80mm]{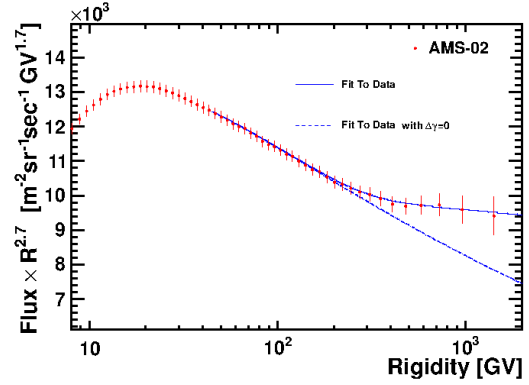} 
\caption{ The AMS proton flux multiplied by $\tilde{R}^{2.7}$ as a function of rigidity $R$.
             The solid curve indicates the fit of Eq.~\ref{eq:fit} to the data. 
             For illustration, the dashed curve uses the same fit values but with 
             $\Delta\gamma$ set to zero.\label{fig:fit}}
\end{figure}

\begin{figure}
\includegraphics[width=80mm]{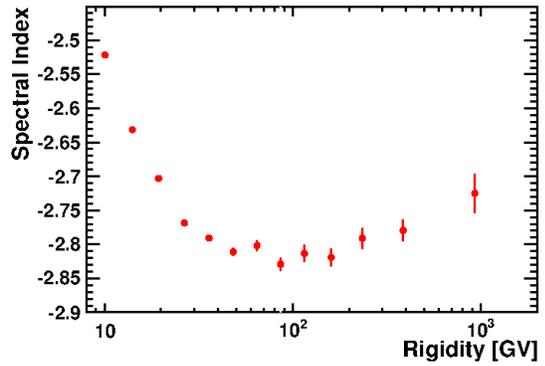} 
\caption{ The dependence of the proton flux spectral index $\gamma$ on rigidity $R$ as obtained from Eq~\ref{eq:gamma}.\label{fig:gamma}}
\end{figure}

The measured proton $\Phi$ flux including statistical
and systematic errors is displayed Figure~\ref{fig:proton} and tabulated in~\cite{bib:table}
as a function of rigidity at the top of the AMS detector.
The contributions to the systematic errors is described in details in~\cite{bib:proton}. 
Figure~\ref{fig:proton} shows the flux as a function of
rigidity with total errors where the rigidity points are placed along the abscissa
at \~{R} calculated for a flux $\propto R^{−2.7}$~\cite{bib:Lafferty}.
A power law $\Phi = C R^{\gamma}$ with a constant spectral index $\gamma$,  
$R$ in GV and $C$ a normalization factor, does not
fit the flux reported in this work and shown in Figure~\ref{fig:proton}
at the 99.9\% C.L. for $R>45$ GV. Applying solar modulation
in the force field approximation \cite{bib:Gleeson} also does not fit
the data at the 99.9\% C.L. for $R>45$ GV. The flux was therefore fitted
with a modified spectral index \cite{bib:Beuermann}
\begin{equation}\label{eq:fit}
\Phi =  C \left(\frac{R}{45\,\mathrm{GV}}\right)^\gamma \left[ 1 +\left(\frac{R}{R_0}\right)^{{\Delta\gamma}/{s}}\right]^s 
\end{equation}
where $s$ quantifies the smoothness of the transition of the spectral index from $\gamma$ 
for rigidities below the characteristic transition rigidity $R_0$ to $\gamma + \Delta\gamma$ 
for rigidities above $R_0$. Fitting over the range 45\,GV to 1.8\,TV yields a $\chi^2/d.f.= 25/26$ 
and parameters values are listened in Table~\ref{table}.
\begin{table}[t]
\begin{center}
\caption{Parameters results from fitting proton flux over the range 45\,GV to 1.8\,TV with minimal model described by Eq~\ref{eq:fit}.}
\begin{tabular}{|c|c|c|c|c|}
\hline \textbf{Fit with Eq.~\ref{eq:fit}} & \textbf{value}   & \textbf{fit-err}    & \textbf{sys-err} & \textbf{sol-err} 
\\
\hline C [$m^{-2}$sr$^{-1}$sec$^{-1}$GV$^{-1}$]         &  0.4544   & $\pm$ 0.0004        & $^{+0.0037}_{-0.0047}$ &  $^{+0.0027}_{-0.0025}$\\
\hline $\gamma$                                         & -2.849    & $\pm$ 0.002         & $^{+0.004}_{-0.003}$   &  $^{+0.004}_{-0.003}$ \\
\hline $\Delta\gamma$                                   & 0.133     & $^{+0.032}_{-0.021}$& $^{+0.046}_{-0.030}$   &  $\pm 0.005$          \\
\hline $s$                                              & 0.024     & $^{+0.020}_{-0.013}$& $^{+0.027}_{-0.016}$   &  $^{+0.006}_{-0.004}$ \\   
\hline $R_0$ [GV]                                       & 336       & $^{+68}_{-44}$      & $^{+66}_{-28}$         &   $\pm 1$             \\
\hline
\end{tabular}
\label{table}
\end{center}
\end{table}
The first error quoted fit-err, takes into account the statistical and uncorrelated systematic errors from
the flux reported in~\cite{bib:proton}. The second sys-err is the error from the remaining systematic errors,
namely from the rigidity resolution function and unfolding, and from the absolute rigidity scale,
with their bin-to-bin correlations accounted for using the migration matrix.
The third sol-err is the uncertainty due to the variation of the solar potential 
$\phi= 0.50$ to 0.62 GV~\cite{bib:modulation}.
The fit confirms that above 45\,GV the flux is incompatible with a single spectral index at the $99.9\%$ C.L.
The fit is shown in Figure~\ref{fig:fit}.
For illustration, the fit result with $\Delta\gamma$ set to zero is also shown in Figure~\ref{fig:fit}.
To obtain the detailed variation of $\gamma$ with rigidity in a model independent way, 
the spectral index is calculated from 
\begin{equation}
\gamma =  d[\log(\Phi)]/d[\log(R)]
\label{eq:gamma}
\end{equation}
over independent rigidity intervals above 8.48\,GV,
with a variable width to have sufficient sensitivity to determine $\gamma$.
The results are presented in Figure~\ref{fig:gamma}. 
The spectral index varies with rigidity  and it is progressively harder with rigidity above $\sim$100\,GV.

\begin{figure}
\includegraphics[width=85mm]{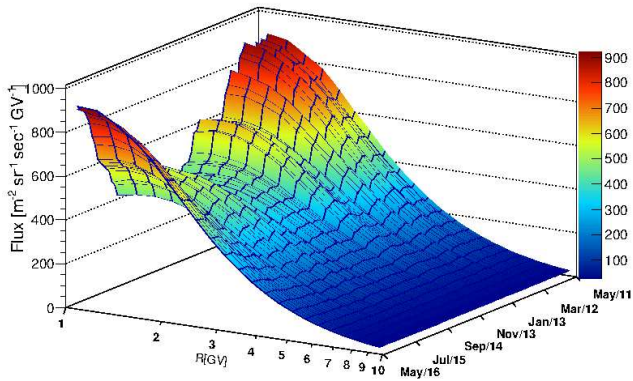} 
\caption{AMS time evolution proton flux from May 2011 to May 2016 in the rigidity range 
from 1 GV to 10 GV.\label{fig:flux3d}}
\end{figure}

While at rigidity $R>45$ GV the spectra remains stable versus time, the low rigidity range
exhibits variations strongly reflecting the solar activity effect.
Figure~\ref{fig:flux3d} displays the proton flux versus time from May 2011 to May 2016 and rigidity from 1 GV to 10 GV.
In the first five years of operations AMS was collecting data during the ascending 
phase of solar cycle 24 through its maximum and toward the minimum.   
The proton spectra was gradually decreasing reaching the minimum between the end of 2013 and the beginning of 2014.
In 2015 the proton flux started to recover: at the end of May 2016, 
AMS proton flux was higher that the one measured at the beginning of the mission.
In addition to the overall modulation effect, the AMS proton flux exhibits structures related to strong solar events 
i.e. Coronal Mass Ejections and Forbush decreases as already reported in~\cite{bib:icrc2015}.

In conclusion, a precision measurement of the proton flux in primary cosmic rays 
with rigidity from 1 GV to 1.8 TV was presented. At high rigidities 
the flux deviates from a single power law and progressively hardens above 100 GV. 
At low rigidities, the proton flux is modulated by the solar activity which reached its maximum in 2014.
In addition to the overall solar modulation effect, the low rigidity flux shows structures 
related to short time scale solar activity that are usually more pronounced during solar maximum. 
Measurement of the time dependent solar modulation of primary cosmic rays fluxes with
AMS on the ISS will be the subject of future publication.

\bigskip 
\begin{acknowledgments}

This work has been supported by acknowledged person and institutions in \cite{bib:proton} and by
National Science Foundation Early Career under grant NSF AGS-1455202, Wyle Laboratories,
Inc. under grant NAS 9-02078, and NASA and Earth
Space Science Fellowship under grant 15-HELIO15F-0005.
\end{acknowledgments}


\bigskip 

\end{document}